\documentclass{ws-ijmpa}

\usepackage{amsfonts,amsmath}
\usepackage{graphicx}
\usepackage{dcolumn}% Align table columns on decimal point
\usepackage{bm}% bold math

\begin{document}

\markboth{D. C. Latimer \& D. J. Ernst}
{Local Demands on Sterile Neutrinos}

%%%%%%%%%%%%%%%%%%%%% Publisher's Area please ignore %%%%%%%%%%%%%%%
%
\catchline{}{}{}{}{}
%
%%%%%%%%%%%%%%%%%%%%%%%%%%%%%%%%%%%%%%%%%%%%%%%%%%%%%%%%%%%%%%%%%%%%

\title{LOCAL DEMANDS ON STERILE NEUTRINOS}

\author{D. C. LATIMER}

\address{
Department of Physics, University of Louisville\\
Louisville, Kentucky  40292, USA \\
dclati01@louisville.edu}

\author{D. J. ERNST}

\address{Department of Physics and Astronomy, Vanderbilt University\\
Nashville, Tennessee 37235, USA\\
david.j.ernst@vanderbilt.edu}

\maketitle

\begin{history}
\received{23 September 2005}
%\revised{Day Month Year}
\end{history}

\begin{abstract}
In a model independent manner, we explore the local implications of a single neutrino 
oscillation measurement which cannot be reconciled within a three-neutrino theory.  We examine this inconsistency for a single region of baseline to neutrino energy $L/E$.
Assuming that sterile neutrinos
account for the anomaly, we find that the {\it local} demands
of this datum can require the addition to the theory of one to three sterile neutrinos. 
We examine the constraints which can be used to determine when more than one 
neutrino would be required. The results apply only to a 
given region of $L/E$. The question of the adequacy of the sterile neutrinos to satisfy a global analysis
is not addressed here.  Finally, using the results of a 3+2 analysis, we indicate values for unknown mixing matrix elements which would require two sterile neutrinos due to local demands only.

\end{abstract}

\ccode{PACS numbers: 14.60.pq}

\section{Introduction}
The experimental evidence for neutrino oscillations is overwhelming 
\cite{solar,superk,lsnd,kamland,k2k,chooz}.  What is more, most of the data 
\cite{solar,superk,kamland,k2k,chooz} can be understood using a three 
neutrino model \cite{3nuanalysis}.  In this, 
the three different flavors of neutrinos are related to three mass eigenstates, 
$\nu_f =U_\mathrm{MNS} \nu_m$, where $U_\mathrm{MNS}$ is the unitary matrix of Maki, 
Nakagawa, and Sakata \cite{mns}.  Recalling the standard parameterization of 
the mixing matrix \cite{pdg}, it is apparent that the parameters in the model 
are three mixing angles, one Dirac CP phase, and two independent mass-squared 
differences.  Solar neutrino experiments \cite{solar} indicate a mass-squared 
difference on the order of $8 \times 10^{-5}$ eV$^2$, and the KamLAND long 
baseline experiment \cite{kamland} confirms this independent of a solar neutrino flux model.
  Studies of atmospheric neutrinos \cite{superk} 
indicate a second mass-squared difference around $2 \times 10^{-3}$ eV$^2$ 
which is consistent with the results from the K2K experiment \cite{k2k}.  The  
neutrino appearance result from the LSND experiment \cite{lsnd} requires a  
mass-squared difference of at least $10^{-1}$ eV$^2$.  With respect to the  
standard three-neutrino theory, these three mass scales are incompatible. 

As most experiments, save LSND, can be explained within the standard  
three-neutrino framework, the LSND result is perplexing;  
however, should the MiniBoone experiment \cite{miniboone} confirm the results  
of LSND, the existence of a third mass-scale will be a reality.  Attempts have
been made to explain all data by the introduction of new physics.
Violation of CPT symmetry for the three neutrinos would allow for additional mass scales \cite{cptv}.
Alternatively, one could introduce into the theory additional neutrinos \cite{sterilenu}.
These additional neutrinos would have to be sterile in order to escape
the limits placed upon the neutrino number by
the invisible decay width of the Z boson.  
The inclusion of one or more sterile neutrinos permits the  
introduction of additional mass scales.  Additionally, one can combine the notions of 
CPT violation and sterile neutrinos in order to attempt to fit the experimental data \cite{sterilecptv}.

We shall consider only the addition of sterile neutrinos.
The simplest extension of the three neutrino model is the inclusion of one  
sterile neutrino.   With four neutrinos, one has three independent mass-squared differences.  
There are two main divisions of mass-squared differences.  One, referred to as the 3+1 scheme, is the 
straightforward extension of the three-neutrino scheme.  The added sterile neutrino is separated 
from the working three-neutrino model with a large mass-squared difference designed to accommodate 
the LSND result.  In the 2+2 scheme, the large mass-squared difference separates two pair of neutrinos 
whose mass-squared differences accommodate the solar and atmospheric data.  Both schemes have at one 
time been promising explanations \cite{fournu}; however, given present data, neither scheme seems to 
provide a compelling description \cite{fournustatus}.  In short, short baseline experiments \cite{chooz,bugey,paloverde,ccfr84,cdhs,karmen,nomad} are 
in tension with the result of the LSND experiment in the 3+1 scheme.  The 2+2 scheme is also disfavored due 
to the results of the solar and atmospheric experiments.  We note that the predictions of the 2+2 scheme are 
slightly better if small, but non-zero, mixing angles are included in the analysis \cite{song}.  As the confidence 
in a four neutrino scenario is underwhelming, five neutrino scenarios have been investigated \cite{sorel,fivenu}.  
It has been shown that the tension among the short-baseline experiments can be relieved somewhat with a 3+2 neutrino 
scheme \cite{sorel}.  
Future experiments could require the addition of even more mass scales and, hence, sterile neutrinos. There is
no theoretical limit on the number that might eventually be required, assuming that cosmological constraints can be avoided. 

In this brief note, we wish to examine in a model independent manner the local implications of a single anomalous neutrino 
oscillation measurement.  
As our concern is local, we shall not consider the whole of the world's data.  Our only appeal to global phenomenology is to 
define a measurement  to be an anomaly if it cannot be reconciled within the standard three-neutrino framework.  
For instance, we class the LSND measurement as anomalous; however, considerations within this paper would not lead 
us to a five neutrino theory because the need for this number of neutrinos comes from a global analysis of data.
Instead, we are interested in whether one can place a local theoretical limit on the effective number of  
sterile neutrinos needed for such an observation.  
In the interest  of parsimony, one would like to have as few additional sterile neutrinos as  
necessary.

\section{Contraction dilation}
We postulate a local region in which the existence of a sterile neutrino(s) is suspected to  
result in an oscillation measurement irreconcilable within a three-neutrino theory.  
As such, we consider a fixed  ratio of baseline to neutrino energy.  We ignore matter effects.  
If indeed a  
sterile neutrino is of consequence in this region, then the MNS matrix which  
relates three mass eigenstates to the three active flavors must be replaced  
by non-unitary mixing relation $T$.
In fact, the  
matrix $T$ can be shown to be a contraction; that is, $T^\dagger T \le 1$, where $1$ is the identity.   
There are an infinite number of ways that $T$ may be dilated into a unitary  
matrix which mixes mass eigenstates and active/sterile neutrinos.  This  
implies that an infinite number of additional neutrinos could be involved in  
creating this anomalous oscillation measurement.  On the other hand, there exists  
a canonical manner by which one may {\it minimally} dilate the contraction  
\cite{sznagy}.  To be definite, let us say that $T$ acts on the space $V$.
We define the defect of the contraction to be 
\begin{equation}
D_T = (1-T^\dagger T)^{1/2}.
\end{equation}
The defect, a positive contraction, indicates, in some sense, how far from  
unitary is $T$.  It can be shown that the following commutation relation  
holds
\begin{equation}
T D_T = D_{T^\dagger} T.
\end{equation}
With this fact, one can prove that the contraction can be minimally dilated to  
the unitary operator
\begin{equation}
U = \left(  
\begin{array}{cc}
T & D_{T^\dagger} \\
D_T & -T^\dagger
\end{array}
\right),
\end{equation}
where $U$ acts upon the larger space $\widehat V := V \oplus
D_{T^\dagger} V$. The dimension  
of this larger space can be calculated 
\begin{equation}
\dim \widehat V = \dim V + \mathrm {rank}\, D_{T^\dagger}.
\end{equation}
The original space $V$ contains three mass eigenstates so its dimension is  
three.  It is thus the rank of the defect which tells us the minimal number  
of mass eigenstates needed {\it locally} at a particular anomalous measurement.  Trivially,  
we have
\begin{equation}
0 \le \mathrm {rank}\, D_{T^\dagger} \le 3.
\end{equation}
If the rank of the defect is zero, then $T$  is actually unitary so that one  
requires no additional sterile neutrinos.  However, for non-zero defect, we see that  
one needs locally at least one sterile neutrino to account  
for the particular anomaly.  In addition, the local demands of the measurement require 
at most three  extra neutrinos.  Obviously, this upper limit is a minimum when 
considering a global analysis of the oscillation data.

This above argument is based upon the non-unitarity of the mixing matrix  
which relates the active flavors of the neutrino with mass eigenstates.  This  
matrix is not directly measurable; rather, it is inferred from measurements.  The inference  
depends upon the number of neutrinos put into the theory from the start,
as such it is model dependent.  
We cannot conclude whether a single anomaly truly does require more  
than one sterile neutrino from the above arguments
alone, for it is conceivable that the contraction $T$ can  
always be chosen in such a manner as to have a defect whose rank is one.   
This would indicate that no more than one sterile neutrino would be necessary to  
explain a single anomaly.  In fact, this is not the case.  We show this now.  

\section{Establishing a lower bound}
Given a neutrino source of flavor $\alpha$, 
one can, at least in principle, measure the flux of flavor $\beta$
neutrinos at the baseline and energy in question. 
We denote this ratio of the measured flux to the
source flux by $\mathcal{P}_{\alpha \beta}$.   We
use lower case Greek letters to indicate the active
neutrino flavors.
We can imagine that all flavors, $\beta = e, \mu,$ and $\tau$, can  
be measured.  For ease of notation, we define the sum over these measured flavors to be
\begin{equation}
c_{\alpha} := \sum_{\beta} \mathcal{P}_{\alpha \beta}.
\end{equation}
Barring exotic physics, the total measured fluxes of active flavors at the
detector must be less than the flux of the source;
hence, the following inequality is satisfied
\begin{equation}
0 \le c_\alpha \le 1. \label{calpha}
\end{equation}
If $c_\alpha$ is unity, then this particular series
of  measurements would not indicate  
any disagreement with respect to a three neutrino theory; otherwise, for this  
quantity less than one, the measurement suggests the need for a sterile  
neutrino or some other physics which we do not consider.  

In principle, one can measure the sum $c_\alpha$
independently for  each active flavor $\alpha= e, \mu,$ and $\tau$ at the  
prescribed ratio of baseline to neutrino energy. This is not 
an experimental reality at the moment; however, in our thought
experiment we are   
permitted such indulgences.  
Each of these measurements satisfy the  
inequality in (\ref{calpha}) independently.  
As we account for the deficit in active flavors by
invoking sterile neutrinos, then the total flux of sterile  
neutrinos at the detector for a given $\alpha$--flavor source is
\begin{equation}
\sum_{j=1}^N \mathcal{P}_{\alpha s_j} = 1 - c_\alpha,
\label{1lessc}
\end{equation}
where the sum is performed over the $N$ sterile
neutrinos.  The index $s_j$ indicates the $j$th
sterile neutrino.  We do not limit the number of
sterile neutrinos {\it a priori}.  

The inclusion of the sterile neutrinos allows us to
treat the ratio of fluxes $\mathcal{P}_{\alpha x}$ as a
probability.  Considering a fixed sterile
neutrino flavor $s_j$ at the detector, we have the
following limit
\begin{equation}
\sum_\alpha \mathcal{P}_{\alpha s_j} \le 1.
\end{equation}
This inequality is valid for each sterile
neutrino.  As such, the double sum over the
three active flavors at the source and all sterile
neutrinos at the detector yields the inequality
\begin{equation}
\sum_{j=1}^N \sum_\alpha \mathcal{P}_{\alpha s_j} \le
N.
\end{equation} 
The expression on the left-hand side of this
inequality can be written in terms of the active
flavor deficits by using Eq.~(\ref{1lessc}); the
result follows easily by summing over the active
flavors at the source
\begin{equation}
\sum_\alpha  \sum_{j=1}^N \mathcal{P}_{\alpha s_j} =
3 -\sum_\alpha c_\alpha.
\end{equation}
Interchanging the sums, we find a lower limit on the
number of sterile neutrinos required to account for a
particular anomaly
\begin{equation}
N \ge 3 - \sum_\alpha c_\alpha.  \label{limit}
\end{equation}
Clearly, if one finds that the sum over the measured active
flavors $c_\alpha$ is less than two,
then one needs more than a single sterile neutrino to
account for this anomaly.

\section{Discussion}
The limit established in Eq.~(\ref{limit}) can be thought of as a means to characterize the degree of the anomaly for a given baseline to neutrino energy ratio.  The burden of information needed to establish the order of an anomaly is extremely high; the oscillation data for all flavors is needed.  In light of this, we consider the application of our treatment to the LSND experimental results in the context of a five neutrino fit of oscillation data.

As stated in the introduction, the results of the LSND experiment \cite{lsnd} cannot be reconciled in the framework of a three neutrino theory with all other existing neutrino oscillation data.  From a muon anti-neutrino beam, the decay at rest (DAR) experiment indicates the appearance of electron anti-neutrinos consistent with a $\overline{\nu}_\mu$--$\overline{\nu}_e$ oscillation probability of $(0.264 \pm 0.081) \%$ at an average baseline to neutrino energy around $1$ m/MeV.  
Three-neutrino analyses neglecting the LSND result provide a compelling explanation of the world's data \cite{3nuanalysis}; however, their prediction of $\mathcal{P}_{\overline{\mu e}}$ at the LSND DAR baseline is on the order of $10^{-7}$.  Hence, we categorize this baseline as one resulting in an anomalous measurement.  Invoking sterile neutrinos as a possible explanation, trivially one must have at least one sterile neutrino.  Of interest, however, is data that requires $N >1$.  The prescription in Eq.~(\ref{limit}) requires knowledge of all active-neutrino oscillation channels at this value for $L/E$, a total of nine independent measurements.  
Assuming that CP symmetry is not violated, then one may reduce the number of needed measurements to six as $\mathcal{P}_{\alpha \beta} = \mathcal{P}_{\beta \alpha}$.  We shall assume that CP is conserved.  

Other than $\overline{\nu}_\mu$--$\overline{\nu}_e$ oscillation probability, there is no other data at the LSND baseline.  There are other short baseline experiments that do have values of $L/E$ in the neighborhood of the LSND measurement which are consistent with no neutrino oscillations and with the three neutrino analyses.  Of these experiments, limitations are placed upon $\overline{\nu}_e$ disappearance by the Bugey \cite{bugey}, CHOOZ \cite{chooz}, and Palo Verde \cite{paloverde} experiments.  The disappearance of $\nu_\mu$ is constrained by the CCFR84 \cite{ccfr84} and CDHS \cite{cdhs} experiments.  Finally, the KARMEN \cite{karmen} and NOMAD \cite{nomad} experiments show no evidence of $\overline{\nu}_\mu$--$\overline{\nu}_e$ and, respectively,  $\nu_\mu$--$\nu_e$ oscillation at values of $L/E$ less than that of the LSND experiment.  At the LSND baseline,  the null-result data of the other experiments indicate that most likely $\mathcal{P}_{ee} \lesssim 1$ and $\mathcal{P}_{\mu \mu} \lesssim 1$.  There is no data in the neighborhood for $\nu_\tau$ channels. 
Given the dearth of data, it is more useful at this point to apply the limit in Eq.~(\ref{limit})  to a neutrino analysis.  

We shall focus upon the 3+2 neutrino analysis presented in Ref.~\refcite{sorel}.  In general, 3+$N$ analyses are tractable models to work with as they are a straightforward extension of three neutrino analyses as in Ref.~\refcite{3nuanalysis}.  For all practical purposes, the three-neutrino fit to the data remains intact while the additional $N$ sterile neutrinos are assumed to have much larger masses which are of consequence for short baselines only.  
Assuming CP is conserved, the probability, {\it in vacuo}, that a neutrino of flavor $\alpha$ and energy $E$ will be detected as a neutrino of flavor $\beta$ a distance $L$ from the source is 
\begin{equation}
\mathcal{P}_{\alpha  \beta}(L/E)
= \delta_{\alpha \beta}
-4 \sum^{3+N}_{\genfrac{}{}{0pt}{}{j >
k}{j,k=1}} U_{\alpha j} U_{\alpha k} U_{\beta k} 
U_{\beta j} \sin^2 (\varphi_{jk})
\label{oscform}
\end{equation}
where $\varphi_{jk} := \Delta_{jk} L/4E$ with $\Delta_{jk} := m_j^2 -
m_k^2$.  
As in Ref.~\refcite{3nuanalysis}, we have mass-squared differences 
$\Delta_{21} \sim 8 \times 10^{-5}$ eV$^2$ and  $\Delta_{31} \sim 2 \times 10^{-3}$ eV$^2$.  Indicated above, these yield oscillation probabilities which are inconsequential for the baselines in the neighborhood of the LSND region so that terms involving these mass-squared differences can be neglected.  Additionally, the mass of each sterile neutrino is assumed to be large in relation to other neutrinos so that at short baselines one may approximate $m_1 \approx m_2 \approx m_3$, or in terms of mass-squared differences $\Delta_{41} \approx \Delta_{42}$, and so on.
Using these approximations and the unitarity of the mixing matrix $U$, one may approximate the oscillation probability (\ref{oscform}) for short baselines given a 3+2 theory
\begin{eqnarray}
\mathcal{P}_{\alpha  \beta}(L/E)
&\approx& \delta_{\alpha \beta}
-4  \, U_{\alpha 5} U_{\alpha 4} U_{\beta 4} 
U_{\beta 5}   \sin^2(\varphi_{54}) \nonumber  \\
&& -4 \left(\delta_{\alpha \beta} - \sum_{j=4}^{5} U_{\alpha j} U_{\beta j} \right) \sum^{5}_{k=4}  U_{\alpha k} U_{\beta k}   \sin^2 (\varphi_{k1}).
\label{approxosc}
\end{eqnarray}
In this case, only two independent mass-squared differences and six mixing matrix elements are relevant.

The 3+2 neutrino analysis presented in Ref.~\refcite{sorel} agrees relatively well with all SBL data, including LSND.  In the context of this particular analysis, we wish to determine whether the LSND result is a first or second order anomaly.  That is, we shall evaluate the limit in Eq.~\ref{limit} at the LSND baseline to determine if the local measurement requires $N > 1$.  
As SBL data concerns only electron and muon neutrino oscillations, analyses can only address four elements of the mixing matrix:  $U_{e4}$, $U_{e5}$, $U_{\mu4}$, and $U_{\mu 5}$. In considering all flavor oscillations, we see from Eq.~\ref{approxosc} that the matrix elements $U_{\tau 4}$ and $U_{\tau 5}$ remain indeterminable.  As such, in evaluating the required number of sterile neutrinos for the local anomaly, these two matrix elements will remain unknown, and we will examine the allowed parameter space for $U_{\tau 4}$ and $U_{\tau 5}$ in order to determine what values, if any, require $N >1$.

The best fit mass-squared differences in the SBL analysis \cite{sorel} are $\Delta_{41}=0.92$ eV$^2$ and $\Delta_{51}=22$ eV$^2$.  Additionally, the best fit mixing matrix elements are $U_{e4}=0.121$, $U_{e5}=0.036$, $U_{\mu 4}=0.204$, and $U_{\mu 5}=0.224$. From Ref.~\refcite{lsnd}, it is apparent that the larger mass-squared differences $\Delta_{51}$ and $\Delta_{54}=\Delta_{51}-\Delta_{41}$ lie outside the resolution of the LSND experiment.    Given this, we shall set $\langle \sin^2(\varphi_{51})\rangle = \langle \sin^2(\varphi_{54})\rangle = 0.5$.  From the best fit parameters, oscillation probabilities at the LSND baseline of 1 m/MeV can be determined from Eq.~\ref{approxosc}:  $\mathcal{P}_{ee}=0.95$, $\mathcal{P}_{\mu \mu}=0.78$, and $\mathcal{P}_{e \mu}=0.29\%$.  In the absence of CP violation, we may simplify the sum
\begin{equation}
\sum_\alpha c_\alpha = \mathcal{P}_{ee}+\mathcal{P}_{\mu \mu}+\mathcal{P}_{\tau \tau} + 2 (\mathcal{P}_{e \mu} +\mathcal{P}_{e\tau} +\mathcal{P}_{\mu \tau}).
\end{equation}
The oscillation probabilities $\mathcal{P}_{\alpha \tau}$ depend upon the unknown matrix elements $U_{\tau 4}$ and $U_{\tau 5}$.  Evaluating the right hand side of Eq.~\ref{limit} at the LSND baseline, we have
\begin{eqnarray}
3 - \sum_\alpha c_\alpha &=& 3.01\, U_{\tau 4}^2 + 1.79\, U_{\tau 5}^2 -0.34\, U_{\tau 4} U_{\tau 5} \nonumber \\
 && -3.39\, U_{\tau 4}^4 -2.00\, U_{\tau 5}^4  -3.39\, U_{\tau 4}^2U_{\tau 5}^2 +0.27. \label{ut4ut5}
\end{eqnarray}
In general, this sum is positive for the allowed values of $U_{\tau 4}$ and $U_{\tau 5}$.  We shall now explore regions in parameter space where this sum is greater than unity, indicating the need for two sterile neutrinos in this region.

Unitarity of the mixing matrix imposes some constraints upon the allowed values of matrix elements $U_{\tau 4}$ and $U_{\tau 5}$.  For one, the following inequality must be satisfied 
\begin{equation}
U_{\tau j}^2  \le 1 - U_{e j}^2 - U_{\mu j}^2, \label{utj}
\end{equation}
for $j = 4,5$.  Employing the best fit parameters from above, we have $\vert U_{\tau 4} \vert \le 0.97$ and $\vert U_{\tau 5} \vert \le  0.97$.  Additionally, unitarity requires
\begin{equation}
U_{\tau 4}^2 + U_{\tau 5}^2  \le 1. \label{circ}
\end{equation}
In fact, the goal of a $3+N$ neutrino analysis is to use the $N$ sterile neutrinos to accommodate the LSND anomaly while keeping the relatively successful three neutrino fit intact; therefore, mixing among active flavor and sterile states should not be too great.  As such, the additional restriction adopted in the 3+2 analysis in Ref.~\refcite{sorel} would limit the unknown matrix elements to $U_{\tau 4}^2 + U_{\tau 5}^2  \le 0.5$.  We shall consider this constraint as well.

\begin{figure}[pt]
\centerline{\psfig{file=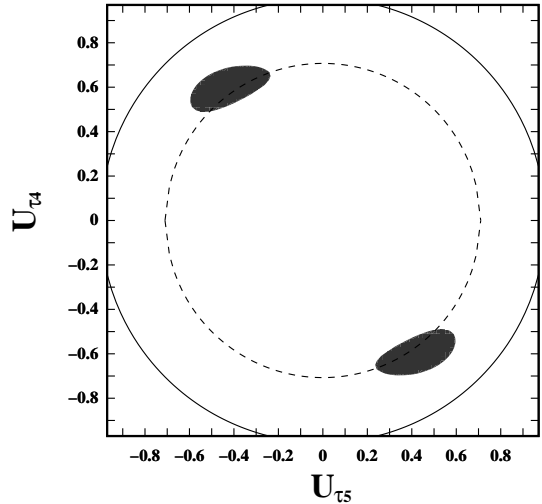,width=7cm}}
\vspace*{8pt}
\caption{ The interior of the solid circle contains the allowed values for the unknown mixing matrix elements $U_{\tau 4}$ and $U_{\tau 5}$.  The shaded regions indicate the values of these parameters which require more than one sterile neutrino for $L/E =$ 1 m/MeV.  The interior of the dashed circle contains the parameters which satisfy $U_{\tau 4}^2 + U_{\tau 5}^2  \le 0.5$\label{fig1}}
\end{figure}

In Fig.~\ref{fig1},  the shaded region in the plot indicates which values of $U_{\tau 4}$ and $U_{\tau 5}$ make the expression in Eq.~\ref{ut4ut5} greater than unity.  The limits on the magnitude of the mixing parameters established in Eq.~\ref{utj} are satisfied.  From the inequality in Eq.~\ref{circ}, the allowed values for the matrix elements lie within the solid circle.  The interior of the dashed circle contains the parameters which satisfy the more rigorous requirement $U_{\tau 4}^2 + U_{\tau 5}^2  \le 0.5$.
From the limit in Eq.~\ref{limit}, the shaded regions in the plot contain matrix elements which {\it require} the existence of two sterile neutrinos at this local measurement.  All other acceptable values of $U_{\tau 4}$ and $U_{\tau 5}$ outside these regions require only one sterile neutrino; as such, the two sterile neutrinos exist merely to satisfy global phenomenological constraints.  

The symmetry of the two disconnected regions in Fig.~\ref{fig1} is no surprise as Eq.~\ref{ut4ut5} shows that measurements of $\mathcal{P}_{\tau \beta}$ oscillations at short baselines can only determine the {\it relative} algebraic sign of $U_{\tau 4}$ and $U_{\tau 5}$.  Perhaps more noteworthy is the fact that a portion of the shaded region lies within the more restrictive dashed circle.   As an example, we shall focus upon one such point:  $U_{\tau 4}=0.6$ and $U_{\tau 5}=-0.3$.   A quick calculation shows that $U_{\tau 4}^2 + U_{\tau 5}^2 =0.45$ and $3- \sum_\alpha c_\alpha = 1.008$.    At the LSND baseline of 1 m/MeV, we may determine the unknown oscillation probabilities for these parameters.  We find $\mathcal{P}_{\tau \tau} = 0.17$, $\mathcal{P}_{\mu \tau} = 0.032$, and $\mathcal{P}_{e \tau} = 0.015$.  Summing up these probabilities over the active flavors, one has $c_\tau = 0.22$.  We recall that if each $c_\alpha$ is unity then no sterile neutrinos are needed locally.  As our choice of parameters require $N > 1$, it is expected that $c_\tau$ should be small, indicating a large coupling of $\nu_\tau$ to the sterile neutrinos.  

\section{Conclusion}
In summation, we began with the assumption that sterile neutrinos
account for an oscillation
datum which is irreconcilable with a three-neutrino
theory.  Given that, we demonstrated that the {\it local} demands
of this datum can require from one to three additional neutrinos.  
We also provide guidance as to how the local measurements might
be used to indicate when more than one neutrino is required.
These additional neutrinos are not the result
of an attempt at a global fit of data; rather, they are needed to account for 
the vanishing of multiple active flavors. 
We made no attempt to determine whether these additional sterile
neutrinos, mandated by local considerations, were
amenable to the global phenomenology of all existing data.  
Such questions were outside the scope of our
initial questions. Global considerations, if not already consistent with 
the local demands, would necessarily
increase the demands beyond that of the single local measurement.  

Given the absence of $\nu_\tau$ oscillation data at short baselines, one cannot use experimental data to evaluate the limit in Eq.~\ref{limit}  in order to determine if the LSND baseline is a region which locally requires more than one sterile neutrino.  However, as an example of practical application, we use the best fit results of a 3+2 neutrino analysis of SBL data \cite{sorel} to determine what values of the mixing matrix elements $U_{\tau 4}$ and $U_{\tau 5}$ would indicate the local need for two sterile neutrinos.  Examining one such point in the parameter space $U_{\tau 4}=0.6$ and $U_{\tau 5}=-0.3$, we see that $\nu_\tau$ would need to be strongly coupled to the sterile neutrinos.  Most likely, this would create tension in the fit of longer baseline data, resulting in the need for additional sterile neutrinos to accommodate global phenomenology.

\bibliography{sterile2b}

\begin{thebibliography}{1}

\bibitem{solar}
B.~T.~Cleveland {\it et al.},
%``Measurement of the solar electron neutrino flux with the Homestake
%chlorine detector,''
Astrophys.\ J.\  {\bf 496}, 505 (1998);
J.~N.~Abdurashitov {\it et al.}  [SAGE Collaboration],
%``Measurement of the solar neutrino capture rate with gallium metal,''
Phys.\ Rev.\ C {\bf 60}, 055801 (1999);
%``Measurement of the solar neutrino capture rate by the Russian-American
%gallium solar neutrino experiment during one half of the 22-year cycle  of
%solar activity,''
J.\ Exp.\ Theor.\ Phys.\  {\bf 95}, 181 (2002)
[Zh.\ Eksp.\ Teor.\ Fiz.\  {\bf 122}, 211 (2002)];
W.~Hampel {\it et al.}  [GALLEX Collaboration],
%``GALLEX solar neutrino observations: Results for GALLEX IV,''
Phys.\ Lett.\ B {\bf 447}, 127 (1999);
M.~Altmann {\it et al.}  [GNO Collaboration],
%``GNO solar neutrino observations: Results for GNO I,''
Phys.\ Lett.\ B {\bf 490}, 16 (2000);
Q.~R.~Ahmad {\it et al.}  [SNO Collaboration],
%``Measurement of the charged current interactions produced by B-8  solar
%neutrinos at the Sudbury Neutrino Observatory,''
Phys.\ Rev.\ Lett.\  {\bf 87}, 071301 (2001);
%``Direct evidence for neutrino flavor transformation from neutral-current
%interactions in the Sudbury Neutrino Observatory,''
Phys.\ Rev.\ Lett.\  {\bf 89}, 011301 (2002);
S.~N.~Ahmed {\it et al.}  [SNO Collaboration],
%``Measurement of the total active B-8 solar neutrino flux at the Sudbury
%Neutrino Observatory with enhanced neutral current sensitivity,''
Phys.\ Rev.\ Lett.\  {\bf 92}, 181301 (2004).

\bibitem{superk}
Y.~Fukuda {\it et al.}  [Super-K Collaboration],
%``Study of the atmospheric neutrino flux in the multi-GeV energy range,''
Phys.\ Lett.\ B {\bf 436}, 33 (1998); 
%``Measurement of the flux and zenith-angle distribution of upward
%through-going muons by Super-Kamiokande,''
Phys.\ Rev.\ Lett.\  {\bf 82}, 2644 (1999); 
S.~Fukuda {\it et al.}  [Super-K Collaboration],
%``Solar B-8 and he p neutrino measurements from 1258 days of
%Super-Kamiokande data,''
Phys.\ Rev.\ Lett.\  {\bf 86}, 5651 (2001);
Y.~Ashie {\it et al.}  [Super-K Collaboration],
%``Evidence for an oscillatory signature in atmospheric neutrino
%oscillation,''
Phys.\ Rev.\ Lett.\  {\bf 93}, 101801 (2004).

\bibitem{lsnd}
C.~Athanassopoulos {\it et al.}  [LSND Collaboration],
%``Evidence for anti-nu/mu $\to$ anti-nu/e oscillation from the LSND
%experiment at the Los Alamos Meson Physics Facility,''
Phys.\ Rev.\ Lett.\  {\bf 77}, 3082 (1996); 
%``Evidence for neutrino oscillations from muon decay at rest,''
Phys.\ Rev.\ C {\bf 54}, 2685 (1996); 
%``Evidence for nu/mu $\to$ nu/e neutrino oscillations from LSND,''
Phys.\ Rev.\ Lett.\  {\bf 81}, 1774 (1998); 
%``Evidence for nu/mu $\to$ nu/e oscillations from pion decay in flight
%neutrinos,''
Phys.\ Rev.\ C {\bf 58}, 2489 (1998);
A.~Aguilar {\it et al.}  [LSND Collaboration],
%``Evidence for neutrino oscillations from the observation of anti-nu/e
%appearance in a anti-nu/mu beam,''
Phys.\ Rev.\ D {\bf 64}, 112007 (2001).

\bibitem{kamland}
K.~Eguchi {\it et al.}  [KamLAND Collaboration],
%``First results from KamLAND: Evidence for reactor anti-neutrino
%disappearance,''
Phys.\ Rev.\ Lett.\  {\bf 90}, 021802 (2003);
T.~Araki {\it et al.}  [KamLAND Collaboration],
%``Measurement of neutrino oscillation with KamLAND: Evidence of spectral
%distortion,''
Phys.\ Rev.\ Lett.\  {\bf 94}, 081801 (2005).

\bibitem{k2k}
M.~H.~Ahn {\it et al.}  [K2K Collaboration],  
%``Indications of neutrino oscillation in a 250-km long-baseline %%@
%experiment,''
 Phys.\ Rev.\ Lett.\  {\bf 90}, 041801 (2003);
%``Search for electron neutrino appearance in a 250-km long-baseline
%experiment,''
Phys.\ Rev.\ Lett.\  {\bf 93}, 051801 (2004);
E.~Aliu {\it et al.}  [K2K Collaboration],
%``Evidence for muon neutrino oscillation in an accelerator-based
%experiment,''
Phys.\ Rev.\ Lett.\  {\bf 94}, 081802 (2005).

\bibitem{chooz}
M.~Apollonio {\it et al.}  [CHOOZ Collaboration],
%``Initial results from the CHOOZ long baseline reactor neutrino  oscillation
%experiment,''
Phys.\ Lett.\ B {\bf 420}, 397 (1998);
%``Limits on neutrino oscillations from the CHOOZ experiment,''
Phys.\ Lett.\ B {\bf 466}, 415 (1999); 
%``Search for neutrino oscillations on a long base-line at the CHOOZ  nuclear
%power station,''
Eur.\ Phys.\ J.\ C {\bf 27}, 331 (2003).

\bibitem{3nuanalysis}
G.~L.~Fogli, E.~Lisi, A.~Marrone and A.~Palazzo,
%``Global analysis of three-flavor neutrino masses and mixings,''
arXiv:hep-ph/0506083;
M.~C.~Gonzalez-Garcia and C.~Pena-Garay,
%``Three-neutrino mixing after the first results from K2K and KamLAND,''
Phys.\ Rev.\ D {\bf 68}, 093003 (2003);
M.~Maltoni, T.~Schwetz, M.~A.~Tortola and J.~W.~F.~Valle,
%``Status of three-neutrino oscillations after the SNO-salt data,''
Phys.\ Rev.\ D {\bf 68}, 113010 (2003).

\bibitem{mns}
Z.~Maki, M.~Nakagawa and S.~Sakata,
%``Remarks On The Unified Model Of Elementary Particles,''
Prog.\ Theor.\ Phys.\  {\bf 28}, 870 (1962).

\bibitem{pdg}
S.~Eidelman {\it et al.}  [Particle Data Group],
%``Review of particle physics,''
Phys.\ Lett.\ B {\bf 592}, 1 (2004).

\bibitem{miniboone}
A.~O.~Bazarko  [BooNe Collaboration],  
%``MiniBooNE: The booster neutrino experiment,'' 
arXiv:hep-ex/9906003.

\bibitem{cptv}
  H.~Murayama and T.~Yanagida,
  %``LSND, SN1987A, and CPT violation,''
  Phys.\ Lett.\ B {\bf 520}, 263 (2001)
  [arXiv:hep-ph/0010178];
  G.~Barenboim, L.~Borissov, J.~Lykken and A.~Y.~Smirnov,
  %``Neutrinos as the messengers of CPT violation,''
  JHEP {\bf 0210}, 001 (2002)
  [arXiv:hep-ph/0108199];
  G.~Barenboim, L.~Borissov and J.~Lykken,
  %``CPT violating neutrinos in the light of KamLAND,''
  arXiv:hep-ph/0212116.

\bibitem{sterilenu}
  J.~T.~Peltoniemi and J.~W.~F.~Valle,
  %``Reconciling dark matter, solar and atmospheric neutrinos,''
  Nucl.\ Phys.\ B {\bf 406}, 409 (1993)
  [arXiv:hep-ph/9302316];
    D.~O.~Caldwell and R.~N.~Mohapatra,
  %``Neutrino mass explanations of solar and atmospheric neutrino deficits  and
  %hot dark matter,''
  Phys.\ Rev.\ D {\bf 48}, 3259 (1993);
  J.~T.~Peltoniemi, D.~Tommasini and J.~W.~F.~Valle,
  %``Reconciling dark matter and solar neutrinos,''
  Phys.\ Lett.\ B {\bf 298}, 383 (1993);
E.~J.~Chun, A.~S.~Joshipura and A.~Y.~Smirnov,
  %``Models of light singlet fermion and neutrino phenomenology,''
  Phys.\ Lett.\ B {\bf 357}, 608 (1995)
  [arXiv:hep-ph/9505275];
  N.~Okada and O.~Yasuda,
  %``A sterile neutrino scenario constrained by experiments and cosmology,''
  Int.\ J.\ Mod.\ Phys.\ A {\bf 12}, 3669 (1997)
  [arXiv:hep-ph/9606411].

\bibitem{sterilecptv}
  V.~Barger, D.~Marfatia and K.~Whisnant,
  %``LSND anomaly from CPT violation in four-neutrino models,''
  Phys.\ Lett.\ B {\bf 576}, 303 (2003)
  [arXiv:hep-ph/0308299].
 
\bibitem{fournu}
    S.~M.~Bilenky, C.~Giunti, W.~Grimus and T.~Schwetz,
  %``Four-neutrino mass spectra and the Super-Kamiokande atmospheric  up-down
  %asymmetry,''
  Phys.\ Rev.\ D {\bf 60}, 073007 (1999)
  [arXiv:hep-ph/9903454];
  V.~D.~Barger, B.~Kayser, J.~Learned, T.~J.~Weiler and K.~Whisnant,
  %``Fate of the sterile neutrino,''
  Phys.\ Lett.\ B {\bf 489}, 345 (2000)
  [arXiv:hep-ph/0008019];
  C.~Giunti and M.~Laveder,
  %``Large nu/mu $\to$ nu/tau and nu/e $\to$ nu/tau transitions in short
  %baseline experiments?,''
  JHEP {\bf 0102}, 001 (2001)
  [arXiv:hep-ph/0010009];
  O.~L.~G.~Peres and A.~Y.~Smirnov,
  %``(3+1) spectrum of neutrino masses: A chance for LSND?,''
  Nucl.\ Phys.\ B {\bf 599}, 3 (2001)
  [arXiv:hep-ph/0011054];
  M.~C.~Gonzalez-Garcia, M.~Maltoni and C.~Pena-Garay,
  %``Solar and atmospheric four-neutrino oscillations,''
  Phys.\ Rev.\ D {\bf 64}, 093001 (2001)
  [arXiv:hep-ph/0105269];
    G.~L.~Fogli, E.~Lisi and A.~Marrone,
  %``Four-neutrino oscillation solutions of the atmospheric neutrino  anomaly,''
  Phys.\ Rev.\ D {\bf 63}, 053008 (2001)
  [arXiv:hep-ph/0009299];
  M.~Maltoni, T.~Schwetz and J.~W.~F.~Valle,
  %``Cornering (3+1) sterile neutrino schemes,''
  Phys.\ Lett.\ B {\bf 518}, 252 (2001)
  [arXiv:hep-ph/0107150].
  
  \bibitem{fournustatus}
    M.~Maltoni, T.~Schwetz and J.~W.~F.~Valle,
  %``Status of four-neutrino mass schemes: A global and unified approach to
  %current neutrino oscillation data,''
  Phys.\ Rev.\ D {\bf 65}, 093004 (2002)
  [arXiv:hep-ph/0112103];   
  M.~Maltoni, T.~Schwetz, M.~A.~Tortola and J.~W.~F.~Valle,
  %``Ruling out four-neutrino oscillation interpretations of the LSND
  %anomaly?,''
  Nucl.\ Phys.\ B {\bf 643}, 321 (2002)
  [arXiv:hep-ph/0207157];
  M.~Maltoni, T.~Schwetz, M.~A.~Tortola and J.~W.~F.~Valle,
  %``Global analysis of neutrino oscillation data in four-neutrino schemes,''
  Nucl.\ Phys.\ Proc.\ Suppl.\  {\bf 114}, 203 (2003)
  [arXiv:hep-ph/0209368];
    M.~Maltoni, T.~Schwetz, M.~A.~Tortola and J.~W.~F.~Valle,
  %``Status of global fits to neutrino oscillations,''
  New J.\ Phys.\  {\bf 6}, 122 (2004)
  [arXiv:hep-ph/0405172].
  
  \bibitem{bugey}
  Y.~Declais {\it et al.},
  %``Search for neutrino oscillations at 15-meters, 40-meters, and 95-meters
  %from a nuclear power reactor at Bugey,''
  Nucl.\ Phys.\ B {\bf 434}, 503 (1995).  

\bibitem{paloverde}
F.~Boehm {\it et al.},
  %``Final results from the Palo Verde neutrino oscillation experiment,''
  Phys.\ Rev.\ D {\bf 64}, 112001 (2001)
  [arXiv:hep-ex/0107009].

\bibitem{ccfr84}
I.~E.~Stockdale {\it et al.},
Phys.\ Rev.\ Lett. {\bf 52}, 1384 (1984).

 \bibitem{cdhs}
  F.~Dydak {\it et al.},
  %``A Search For Muon-Neutrino Oscillations In The Delta M**2 Range 0.3-Ev**2
  %To 90-Ev**2,''
  Phys.\  Lett.\  B {\bf 134}, 281 (1984).

\bibitem{karmen}  
B.~Armbruster {\it et al.}  [KARMEN Collaboration],
  %``Upper limits for neutrino oscillations anti-nu/mu $\to$ anti-nu/e from
  %muon decay at rest,''
  Phys.\ Rev.\ D {\bf 65}, 112001 (2002)
  [arXiv:hep-ex/0203021].
  
\bibitem{nomad}
  P.~Astier {\it et al.}, 
  Phys. Lett. {\bf B 570}, 19 (2003).
  
  
\bibitem{song}
H.~Pas, L.~g.~Song and T.~J.~Weiler,
  %``The hidden sterile neutrino and the (2+2) sum rule,''
  Phys.\ Rev.\ D {\bf 67}, 073019 (2003)
  [arXiv:hep-ph/0209373].

\bibitem{sorel}
  M.~Sorel, J.~M.~Conrad and M.~Shaevitz,
  %``A combined analysis of short-baseline neutrino experiments in the (3+1)
  %and (3+2) sterile neutrino oscillation hypotheses,''
  Phys.\ Rev.\ D {\bf 70}, 073004 (2004)
  [arXiv:hep-ph/0305255].
  
  \bibitem{fivenu}
  K.~S.~Babu and G.~Seidl,
  %``Simple model for (3+2) neutrino oscillations,''
  Phys.\ Lett.\ B {\bf 591}, 127 (2004)
  [arXiv:hep-ph/0312285];
 K.~L.~McDonald, B.~H.~J.~McKellar and A.~Mastrano,
  %``(3+2) neutrino scheme from a singular double see-saw mechanism,''
  Phys.\ Rev.\ D {\bf 70}, 053012 (2004)
  [arXiv:hep-ph/0401241];  
  W.~Krolikowski,
  %``Two light sterile neutrinos that mix maximally with each other and
  %moderately with three active neutrinos,''
  Acta Phys.\ Polon.\ B {\bf 35}, 1675 (2004)
  [arXiv:hep-ph/0402183].

\bibitem{sznagy}
B. Sz.-Nagy and C. Foias, {\it Harmonic Analysis of Operators on Hilbert 
Space} (North-Holland Pub. Co., Amsterdam, 1970).


\end{thebibliography}

\end{document}